\documentclass[conference]{IEEEtran}
\IEEEoverridecommandlockouts
\usepackage{cite}
\usepackage{amsmath,amssymb,amsfonts}
\usepackage{algorithm}
\usepackage{algpseudocode}
\usepackage{graphicx}
\usepackage{textcomp}
\usepackage{xcolor}
\usepackage{multirow}
\usepackage{multicol}
\usepackage{booktabs}
\usepackage{float}
\usepackage{makecell}
\usepackage{caption}
\usepackage{lipsum}
\usepackage{amssymb}
\usepackage{pifont}
\usepackage{url}
\include {scalerel}
\makeatletter
\newcommand{\linebreakand}{%
  \end{@IEEEauthorhalign}
  \hfill\mbox{}\par
  \mbox{}\hfill
  \begin{@IEEEauthorhalign}
}
\makeatother

\def\BibTeX{{\rm B\kern-.05em{\sc i\kern-.025em b}\kern-.08em
    T\kern-.1667em\lower.7ex\hbox{E}\kern-.125emX}}
\begin{document}

\title{Understanding Data Movement in Tightly Coupled Heterogeneous Systems: A Case Study with the Grace Hopper Superchip\\
{}
}

\author{\IEEEauthorblockN{Luigi Fusco}
\IEEEauthorblockA{\textit{Department of Computer Science} \\
\textit{ETH Zurich}\\
Zurich, Switzerland \\
luigi.fusco@inf.ethz.ch}
\and
\IEEEauthorblockN{Mikhail Khalilov}
\IEEEauthorblockA{\textit{Department of Computer Science} \\
\textit{ETH Zurich}\\
Zurich, Switzerland \\
mikhail.khalilov@inf.ethz.ch}
\and
\IEEEauthorblockN{Marcin Chrapek}
\IEEEauthorblockA{\textit{Department of Computer Science} \\
\textit{ETH Zurich}\\
Zurich, Switzerland \\
marcin.chrapek@inf.ethz.ch}
\linebreakand
\IEEEauthorblockN{Giridhar Chukkapalli}
\IEEEauthorblockA{
\textit{NVIDIA}\\
Santa Clara, California \\
gchukkapalli@nvidia.com}
\and
\IEEEauthorblockN{Thomas Schulthess}
\IEEEauthorblockA{\textit{Swiss National Supercomputing Centre (CSCS)} \\
\textit{ETH Zurich}\\
Zurich, Switzerland \\
schulthess@cscs.ch}
\and
\IEEEauthorblockN{Torsten Hoefler}
\IEEEauthorblockA{\textit{Department of Computer Science} \\
\textit{ETH Zurich}\\
Zurich, Switzerland \\
torsten.hoefler@inf.ethz.ch}
}
%\maketitle

\twocolumn[{
\renewcommand\twocolumn[1][]{#1}
\maketitle
\vspace{-0.5cm}
\begin{center}
\captionsetup{type=figure}
\includegraphics[width=\textwidth]{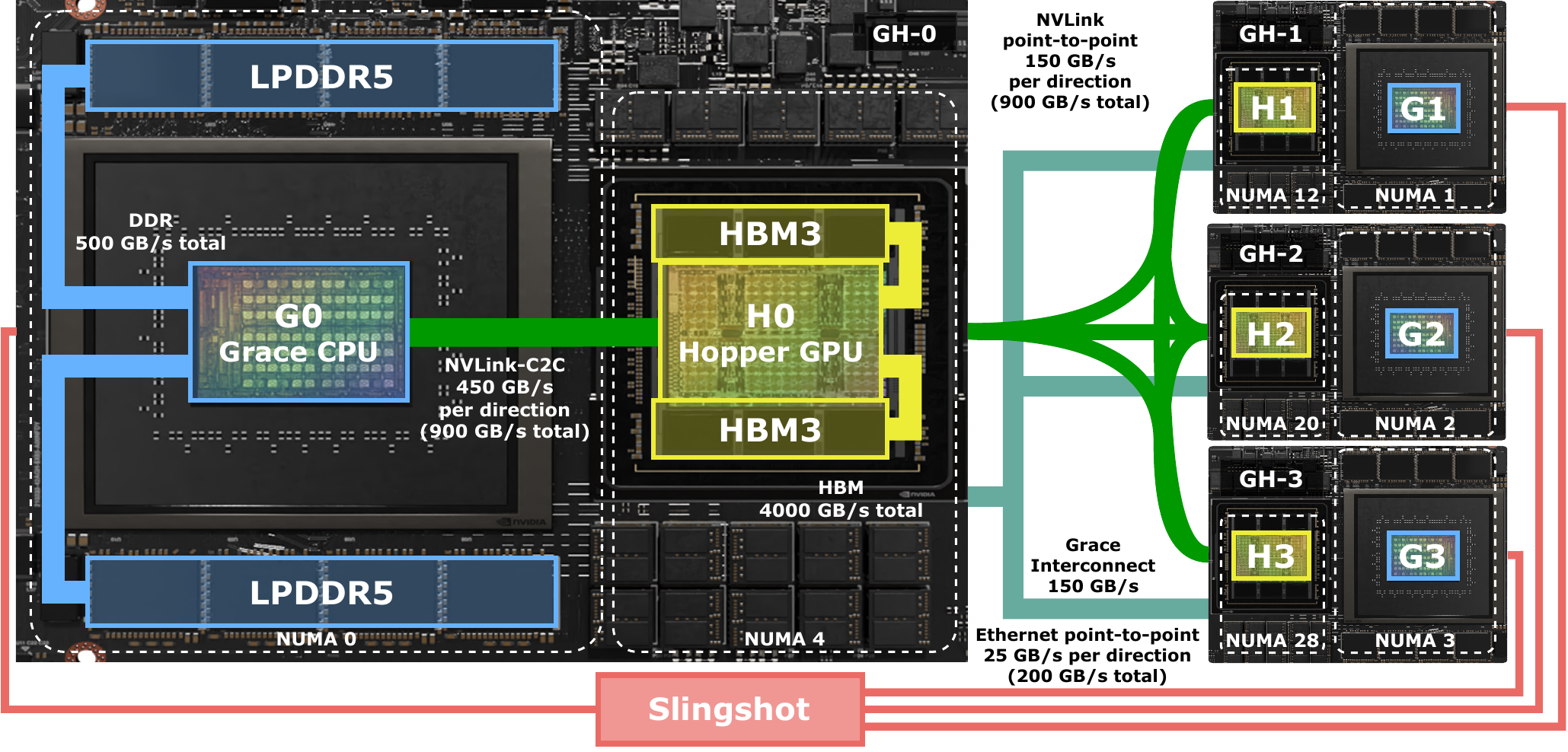}

\captionof{figure}{Architecture of the Quad GH200 node of the Alps supercomputer. Every node is composed of four GH200 fully connected using NVLink and a cache coherent interconnect. Every GH200 is connected to a Slingshot network through a separate NIC.}
\label{full}
\end{center}%
\vspace{0.2cm}
}]

\begin{abstract}
Heterogeneous supercomputers have become the standard in HPC. GPUs in particular have dominated the accelerator landscape, offering unprecedented performance in parallel workloads and unlocking new possibilities in fields like AI and climate modeling.
With many workloads becoming memory-bound, improving the communication latency and bandwidth within the system has become a main driver in the development of new architectures.
The Grace Hopper Superchip (GH200) is a significant step in the direction of tightly coupled heterogeneous systems, in which all CPUs and GPUs share a unified address space and support transparent fine grained access to all main memory on the system.
We characterize both intra- and inter-node memory operations on the Quad GH200 nodes of the new Swiss National Supercomputing Centre Alps supercomputer, and show the importance of careful memory placement on example workloads, highlighting tradeoffs and opportunities.
\end{abstract}


\section{Introduction}
Heterogeneous platforms are dominant in modern-day large-scale computing. GPUs are ubiquitous in the fields of HPC and AI and have permitted the growth of workloads to unprecedented scales \cite{matsuoka2009gpu,navarro2014survey,giorgetta2022icon}. Recent breakthroughs in generative AI are made possible by the availability of computational resources, with the need for memory and computation growing steadily. With Large Language Models (LLMs) breaking the trillion parameter size, memory is a critical resource that enables large-scale training and inference \cite{isaev2023scaling}.

GPUs were born as standalone accelerators. An application is run as a program on the CPU, which is responsible for orchestrating the execution of computational kernels on the GPU and for managing memory allocations and data transfers. Data transfers are critical for the performance of applications regardless of the access pattern, that can range from frequent low-latency communications to bulk copies \cite{kato2013zero,van2014performance,bauer2011cudadma}.
The importance of optimizing data movement has been reflected in the development of data allocation and management APIs, unified memory systems, more advanced interconnects, and new programming models \cite{li2019evaluating,zhang2023perks}. With the NVLink-C2C (C2C) interconnect, allowing for fast, low latency, cache coherent interaction between different classes of chiplets, NVIDIA has marked the beginning of a new class of high-end tightly coupled systems where every Processing Unit (PU) has complete access to all main memory on the system through a unified memory space.

NVIDIA’s GH200 Grace Hopper Superchip (GH200) connects an ARM CPU and a Hopper GPU through the C2C interconnect. Multiple GH200s can be connected to create a large-scale tightly coupled heterogeneous system. We explore the performance characteristics of the Quad GH200 system, the building block of the new Swiss National Supercomputing Centre Alps supercomputer, through a series of microbenchmarks. Many works already look at benchmarking heterogeneous systems and how to program them \cite{mittal2015survey}, studying the performance characteristics of heterogeneous workloads \cite{shen2013glinda,shen2014look,shen2015workload,mistry2013valar,stratton2012parboil} or executing simple parallel programs on different architectures \cite{danalis2010scalable,che2009rodinia}.

Some works already looked at the capabilities of the GH200 shared memory system, offloading automatically BLAS kernels to the GPU in CPU-only scientific codes \cite{li2024automatic} and studying the effect of different memory allocation policies on different applications \cite{schieffer2024harnessing}.

The objective of our microbenchmarks is to analyze the interaction between all PUs, physical memories, and memory allocation APIs of the system, highlighting tradeoffs and opportunities. Growth in computing power was not matched by improvements in memory access latency and bandwidth, resulting in many applications becoming memory-bound. Data movement is now the dominant factor for HPC and ML workloads, and its optimization is crucial to application performance \cite{mccalpin1995memory,unat2017trends,ben2019stateful,ivanov2021data}.

Tightly coupled systems, like the Quad GH200, greatly expand the design space for these optimizations, allowing for a much larger choice when deciding where to place data and compute. Having access to a larger pool of memory opens up new possibilities for scaling applications with large memory footprints that go beyond what is directly available to a single GPU or CPU. As this pool is heterogeneous, informed data placement is of crucial importance.

We make the following contributions:
\begin{itemize}
    \item We design a comprehensive set of microbenchmarks to analyze memory operations in complex tightly coupled heterogeneous systems\footnote{code at \texttt{https://github.com/luigifusco/gh\_benchmark}}.
    \item We validate our datapath-oriented approach by highlighting the effect on the performance of sample workloads of different data placement policies.
    \item We present a comprehensive analysis of the Quad GH200 node of the Alps supercomputer.
\end{itemize}
\section{Background}
We describe the Grace Hopper Superchip and the architecture of the tested system (Figure \ref{full}) with a bottom-up approach, starting from the description of its fundamental hardware components up to the software level, focusing on memory subsystems and management.
\subsection{The Architecture}
The NVIDIA GH200 Grace Hopper Superchip (GH200) is a heterogeneous coherent system that combines a Grace CPU and a Hopper GPU. For the rest of the paper, we refer to any of these two types of chips as Processing Units (PU).
\subsubsection{NVLink-C2C}
NVLink is an interconnect technology originally designed by NVIDIA as an alternative to PCIe providing additional features and targeting multi-GPU systems. It evolved through generations improving link speeds, and has reached its fourth iteration. NVLink-C2C (C2C) extends the NVLink family with a high-speed interconnect to engineer integrated devices built by combining multiple chiplets. It allows fast and cache coherent communication between different classes of PUs. Its architecture allows for a bandwidth of 40 Gbps for every data signal, with every link supporting 9 data signals. In a GH200, Grace and Hopper incorporate ten links each, for a total bandwidth of 450 GB/s per direction \cite{wei20239}.

C2C supports the Arm AMBA Coherent Hub Interface (AMBA CHI) architecture, which defines a scalable and coherent hub interface and on-chip interconnect \cite{ambachi}. AMBA CHI allows for modular designs. It supports cache coherency at the 64-byte granularity, snoop filtering, and different cache models with data forwarding, atomic operations and synchronization, and virtual memory management.

\subsubsection{Grace}
Grace is an HPC-oriented CPU incorporating 72 Arm Neoverse V2 CPU cores, a 64-bit data center-oriented architecture. It supports up to 480GB of LPDDR5 ECC memory with a bandwidth of up to 500GB/s. The CPU cores are distributed throughout the Scalable Coherency Fabric, a mesh fabric providing up to 3.2TB/s of total bisection bandwidth, integrating CPU cores, memory, system IOs, and C2C connections. Each core has 64KB of L1 instruction cache, 64KB of L1 data cache, and 1MB of L2 data cache. The size of the shared L3 data cache is 114MB. The cache coherency protocol is MESI with inclusive L2 cache \cite{neoverse}.% Grace can be found in the GH200 but is also the main component of the Grace Superchip, which connects two Grace CPUs with the C2C interconnect into a system with a total of 144 cores and 234MB of shared L3 cache.

\subsubsection{Hopper}
\begin{table}
\centering
\caption{Main hardware features of the available GPUs implementing the Hopper architecture.}
\begin{tabular}{c | c c c c}
    \toprule 
    \textbf{Feature} & \multicolumn{2}{c}{\textbf{GH200}} & \textbf{H100 SXM} & \textbf{H100 PCIe} \\
    \midrule
    SMs & \multicolumn{2}{c}{132} & 132 & 114 \\
    Memory Type & HBM3 & HBM3e & HBM3 & HBM2e \\
    Memory Size & 96GB & 144GB & 80GB & 80GB \\
    Bandwidth & 4TB/s & 4.9TB/s & 3.35TB/s & 2TB/s \\
    \bottomrule
\end{tabular}
\label{hoppers}
\end{table}
The Hopper architecture, the successor of the Ampere architecture, was launched by NVIDIA in 2022. It is designed for data center use and is parallel to the consumer-oriented Ada Lovelace architecture. It is currently employed in the variants of the H100 GPU and GH200. Table \ref{hoppers} lists the main hardware characteristics of these GPUs. Two variants of  Hopper in GH200 exist. We have access to the one with 132 Streaming Multiprocessors (SMs) and connected to 96GB of HBM memory with a bandwidth of over 4TB/s. Each SM has 256KB of private L1 cache and can run up to 2048 concurrent threads, for a maximum total of 270,336 concurrent threads. 52MB of L2 cache are shared between all SMs. It supports up to 18 NVLink 4 connections at 25 GB/s per direction each, for a maximum bandwidth of 450 GB/s per direction.

\subsubsection{Grace Hopper}
The C2C interconnect provides cache-coherent memory access between the Grace CPU and the Hopper GPU on a GH200. The same technology is used to connect the two Grace PUs in the Grace Superchip, combining two Grace CPUs, and will be used in the recently announced Grace Blackwell Superchip, combining a Grace CPU and two Blackwell GPUs. The interconnect provides a bidirectional bandwidth of 900GB/s, 7x higher than what is achievable by the H100 which uses PCIe 5 with a bidirectional bandwidth of 128 GB/s. Up to 32 GH200 can be connected through the NVIDIA NVLink Switch System. All interconnected GH200 will act as a single cache coherent system. This allows all Hoppers to communicate with with each other at a bidirectional bandwidth of 900 GB/s, for a total of 19.5 TB of shared memory in a single cache coherent system supporting direct load, stores, and atomic operations. A single shared memory system is also provided in the Quad GH200 configuration, composed of four fully interconnected Superchips through NVLink. This configuration offers a lower interconnection bandwidth, as the 18 links that the Hopper architecture provides are equally split between three channels. In the rest of the paper, we refer to this system as being composed by \textit{peer} GH200s.

\subsubsection{Alps Supercomputer}
% \begin{figure}
% \centering
% \includegraphics[width=\linewidth]{images/blade.png}
% \caption{Picture of a HPE Cray Supercomputing EX254n blade, hosting two nodes, four GH200 each.}
% \label{blade}
% \end{figure}
For our analysis, we use the early access Santis partition of the Alps supercomputer developed by the Swiss National Supercomputing Centre (CSCS), which is currently under provisioning. The system is made up of HPE Cray Supercomputing EX254n blades, each hosting two nodes. Figure \ref{full} provides an overview of a node\footnote{GH200 image source: \url{https://developer-blogs.nvidia.com/wp-content/uploads/2022/08/image3-8.png}}. Each node is composed of four interconnected GH200. Every Superchip is connected to every other through NVLink and a cache coherent interconnect, from now on defined as Grace Interconnect (GI). Grace traffic to peer Superchips is routed through the GI interconnect, while Hopper traffic is routed through NVLink. 

Every GH200 has 96 GB of HBM3 and 128 GB of LPDDR5 memory, for 896 GB of total memory in its final configuration\footnote{at time of writing, only 120 GB of LPDDR5 memory per GH200 were available}. Every Quad GH200 node acts as a single NUMA system, with 288 CPU cores and 4 GPUs. Nodes are interconnected using HPE Slingshot 11 \cite{de2020depth} in a dragonfly topology \cite{kim2008technology}, with 4 injection ports per node. Each NIC is connected to a separate GH200 through a mezzanine card with two PCIe x16 Gen5 connections, for a total of 32 lanes. Each GH200 is connected to a switch through a 200 Gb/s ethernet port, for a total of 4 ports per node, and a maximum achievable bidirectional bandwidth of 100 GB/s to other nodes.
The system page size is configured to be 64 KB.
%The Grace revision is A02, with SoC ID \texttt{0x0241} and SoC Revision \texttt{0x00000102} according to the JEDEC JEP-106 Manufacturer ID Code.
The clock speed is 3,483 MHz without frequency boost. At the time of writing the system runs a software stack that will change once in production. The nodes run SUSE Linux Enterprise Server 15 SP5. The NVIDIA driver is version 535.129.03. We use CUDA 12.3 and GCC 12.3. 

\subsection{Memory Hierarchy}
\begin{figure}
\centering
\includegraphics[width=\linewidth]{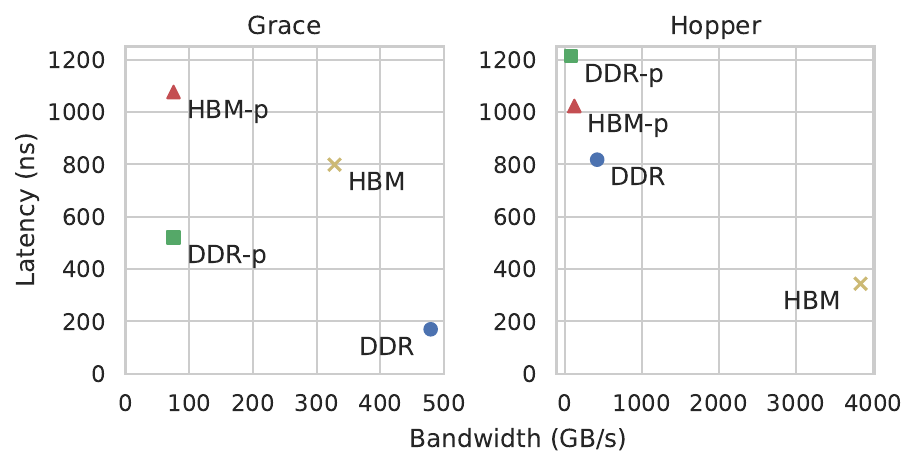}
\caption{Maximum bandwidth plotted against access latency achieved by Grace (left) and Hopper (right) to different memories of the system. The suffix "-p" indicates memory on a peer GH200.}
\label{bandwidth_latency_plot}
\end{figure}

\begin{figure*}[!t]
\centering
\includegraphics[width=.9\textwidth]{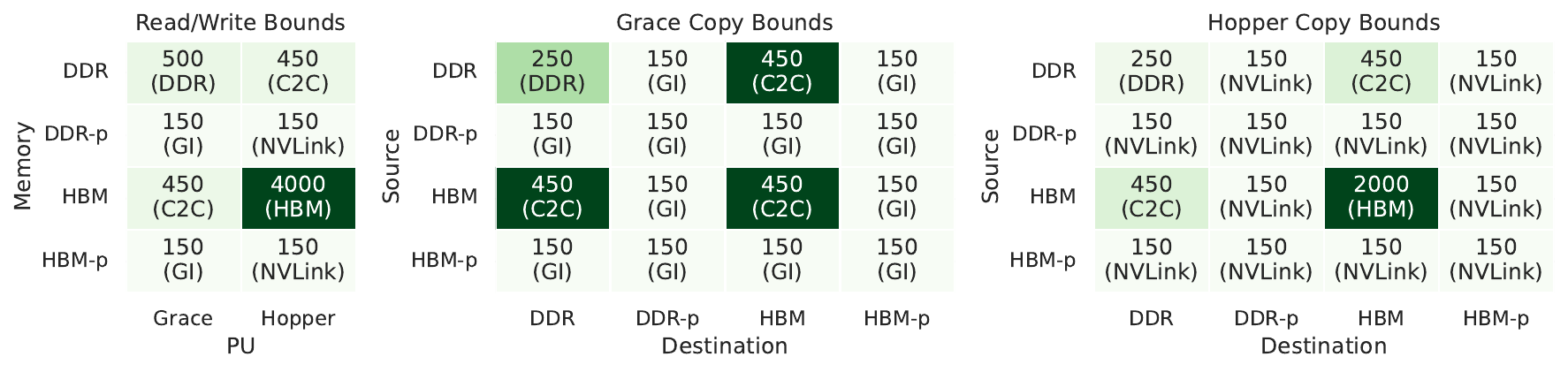}
\caption{Theoretical bandwidth bound for (left to right) read and write operations, copy operations issued by a Grace, and copy operations issued by a Hopper. Bounds depend on the datapath, which in turn depends on the types of memories involved. The bounds are shown in GB/s and include the limiting interconnect.}
\label{bounds}
\end{figure*}

The performance of modern computing systems is often limited by memory access speed. Processors have evolved to include larger and more complex cache hierarchies, with faster access to memories close to where the computation is performed. This hierarchical approach is also applied when designing main memory systems, fueled by a growth in the number of processors to serve. In these complex systems, the datapath depends on which physical memory bank is accessed. Compute units have direct access to some memory controllers, but need to send requests through an interconnect to access others.
This adds complexity when designing applications, as complex interactions between PUs and memories emerge \cite{brecht1993importance,ramos2017capability,milic2017beyond}.

Non-Uniform Memory Access (NUMA) is a logical division of memory supported by modern operating systems. It consists of defining the affinity of cores to different regions of memory. 
%Typical server architectures have multiple NUMA nodes, each composed of a processor, and joined by an interconnect. Memory requests to remote nodes go through the interconnect, resulting in slower access due to intermediate protocols, longer physical distances, and possible congestion issues.
%Modern operating systems and applications are designed to be NUMA-aware, optimizing their operations by allocating resources and scheduling processes in a way that minimizes cross-node memory access and takes advantage of the local memory access speeds.
Every GH200 is composed of two NUMA nodes. One consists of the LPDDR5 memory with affinity to Grace, while the other consists of the HBM3 memory with affinity to Hopper. Interconnected Superchips in a Quad GH200 appear as a single device with two NUMA nodes per unit. Figure \ref{full} highlights the different NUMA nodes that compose a node in the Alps supercomputer. The four Grace CPUs are associated with NUMA nodes 0, 1, 2, and 3, following a sequential numbering. The four Hopper GPUs are associated with NUMA nodes 4, 12, 20, and 28.

Physical allocation on NUMA nodes follows a first access principle for memory allocated through system calls like \texttt{brk} or \texttt{mmap}, and for APIs like \texttt{malloc} and \texttt{new}. Alternatively, the NUMA node can be explicitly chosen by using the numactl utility command, through libnuma,  or using \texttt{numa\_alloc\_onnode}, which lets the user specify the id of the NUMA node where to allocate memory.

\subsection{Unified Memory}
\begin{table*}[h]
\caption{Types of memory and relative API available on GH200.}
\centering
\begin{tabular}{c c c c c c}
    \toprule
    \textbf{Type} & \textbf{API} & \textbf{Placement} & \textbf{GPU Translation} & \textbf{Page Size} & \textbf{Automatic Migration} \\
    \midrule
    \multirow{ 2}{*}{System-allocated} & \texttt{mmap}, \texttt{malloc}, \texttt{new} & First touch & \multirow{ 2}{*}{ATS} & \multirow{ 2}{*}{System} & Yes (CUDA\(\geq\)12.4) \\
    \addlinespace[0.4em]
     & \texttt{numa\_alloc\_onnode} & Specified & & & No \\
    \addlinespace[0.6em]
    Device & \texttt{cudaMalloc} & HBM & GPU-MMU & 2 MB & No \\
    \addlinespace[0.5em]
    Managed & \texttt{cudaMallocManaged} & First touch & \makecell{ATS (DDR)\\ GPU-MMU (HBM)} & \makecell{System (DDR)\\ 2 MB (HBM)} & Yes \\
    \addlinespace[0.4em]
    Pinned & \texttt{cudaMallocHost}, \texttt{cudaHostAlloc} & DDR & ATS & System & No \\
    \bottomrule
\end{tabular}
\label{memory_types}
\end{table*}

\begin{figure}[h]
\centering
\includegraphics[width=\linewidth]{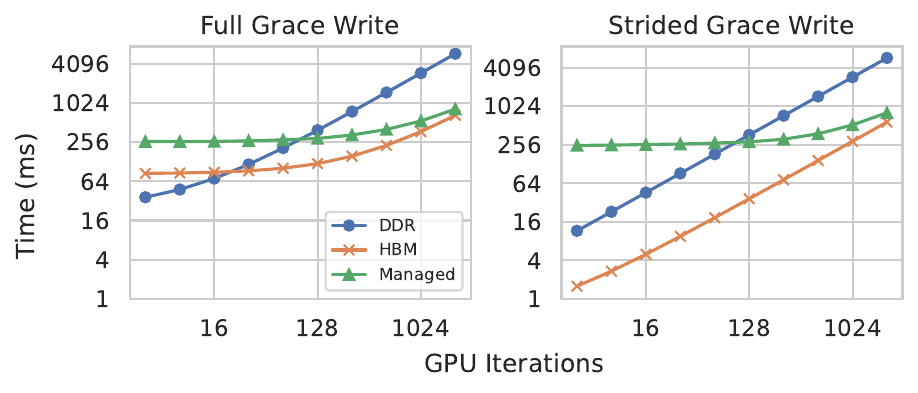}
\caption{Execution time of an artificial application using system-allocated memory and managed memory (lower is better).}
\label{managed_app}
\end{figure}

Heterogeneous systems typically require fine-grained control of memory, with research focusing on facilitating its management \cite{dashti2017analyzing,wang2018superneurons,haria2018devirtualizing}. Memory allocations are device-specific and live in separate memory spaces. Memory allocated using \texttt{cudaMalloc} can be accessed only by the GPU, and memory transfers between the GPU and CPU address space require the utilization of specific copy APIs like \texttt{cudaMemcpy}.

Unified Virtual Addressing (UVA) simplifies this programming model by enabling devices (GPUs) to share a unified address space with the host (CPU), allowing every PU on the system to access memory using the same pointers. It was first introduced with CUDA 4 and allowed easy access to peer GPU memory in a multi-GPU system and zero-copy access to host pinned memory allocated using \texttt{cudaMallocHost} or \texttt{cudaHostAlloc} through DMA over PCIe. Managed Memory was introduced with CUDA 6. By calling \texttt{cudaMallocManaged} the user can obtain the pointer to a Managed Memory region that is accessible by both the CPU and the GPU through automatic copies. In Kepler architectures calling \texttt{cudaMallocManaged} results in memory being allocated on the device. As the CPU accesses this memory a page fault is triggered and the CUDA driver migrates the page from the device to the host. On kernel launch, all managed memory is migrated to the device. Pascal introduced the support for device page faults and migrations, removing the need for all memory to be copied on the device on kernel launch and introducing a first touch policy for the physical allocation of pages.

In a Linux system the page table stores virtual to physical address translations. These translations are cached in a translation-lookaside buffer (TLB) for faster access. The Memory Management Unit (MMU) is responsible for performing these translations. In a traditional system, the CPU and GPU have distinct MMUs and TLBs and work on separate virtual and physical addresses using separate page tables. Address Translation Services (ATS) extend the PCIe protocol to support caching of address translations. A miss in the device MMU will result in an Address Translation Request to the CPU. The CPU checks its page tables for the virtual-to-physical mapping for that address and supplies the translation back to the GPU, which will store it in its local Address Translation Cache \cite{armats}. Address Translation Services (ATS) were introduced in CUDA 9.2 for integration in IBM Power 9 systems using Volta GPUs through NVLink connections, and allowed for fine-grained access to memory, serving loads and stores at the cache line level \cite{gayatri2019comparing}.

In a Grace Hopper system, a single page table and virtual address space are shared between CPU and GPU. A specific unit called Address Translation Service Translation Buffer Unit (ATS-TBU) is implemented to provide fast translations and support interaction between all MMUs and TLBs on the system. Managed memory, on the other hand, requires a full page transfer if memory is not local on the PU.

Managed memory was already shown to outperform system memory accessed through ATS in applications with frequent memory access performed by the GPU \cite{gayatri2019comparing}. To give an idea of the difference between the two systems we develop a simple application that interleaves a series of back-to-back Hopper-issued writes with Grace-issued writes. Hopper writes use \texttt{cudaMemset} while Grace writes use either \texttt{memset} or strided stores. By setting the stride to the page size of 64 KB we test the worst use case for managed memory, with the least number of bytes used per bytes transferred.

Figure \ref{managed_app} shows the time it takes to execute the application with different numbers of iterations and different types of allocations (lower is better). The runtime of managed memory gets asymptotically close to system-allocated memory on HBM. Managed memory results in a faster runtime only after a large number (at the 128 mark) of back-to-back iterations on Hopper. For a large number of Hopper iterations, there is no difference between the full write and strided write version, showing that the workload is heavily GPU-bound. For a low number of iterations, the full write version shows better performance on DDR than on HBM, showing that the workload is CPU-bound. ATS shines in workloads where the access patterns are more complex than sequential reads and writes.

%Another recent addition to the shared memory landscape is Heterogeneous Memory Management (HMM). HMM is a Linux kernel subsystem that provides an infrastructure to integrate heterogeneous systems by creating a single unified address space and has been supported since CUDA 12.2. HMM works by offering an API to easily update device page tables and keep them synchronized with the CPU page tables, as well as transferring memory by using existing page migration mechanisms. With HMM enabled, GPUs can access all system-allocated memory without any explicit management.

%CUDA supports memory allocation and management only on the host. The host must allocate and prepare memory before a device kernel is executed. 

Table \ref{memory_types} provides an overview of the different types of memory that can be allocated on the GH200, describing the available APIs and the physical placement of the allocated memory. System-allocated memory is made accessible by both host and device through ATS, and benefits from fine-grained access through the C2C interconnect. \texttt{cudaMalloc} allocates memory on the GPU. It is the only listed API that does not allow direct CPU access. \texttt{cudaMallocHost} allocates pinned memory on the host, allowing for streaming access from the device through DMA engines. This is necessary for \texttt{cudaMemcpyAsync} to work asynchronously, and can improve CPU-GPU memory transfer speeds \cite{pearson2019evaluating}. \texttt{cudaMallocManaged} allocates managed memory in a uniform memory space shared between host and device and is managed by the NVIDIA driver at a page level.

\begin{figure}
\includegraphics[width=\linewidth]{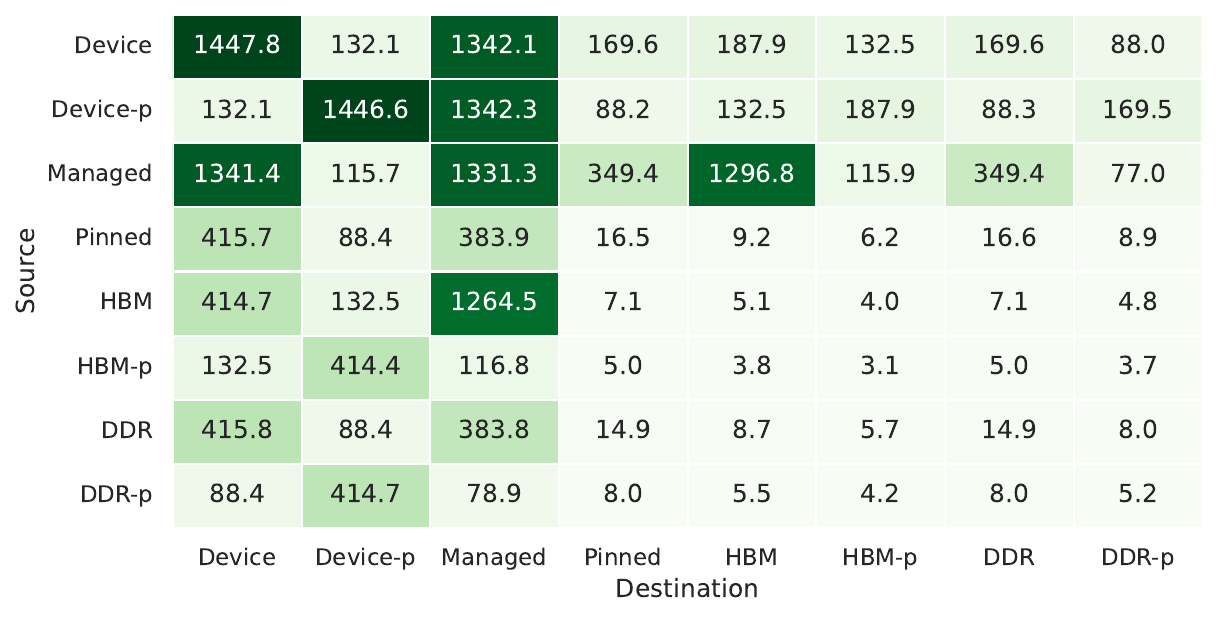}
\caption{Throughput (GB/s) achieved by cudaMemcpy on different combinations of source and destination memory types. Note that \textit{Device} and \textit{HBM} memory are physically co-located but are allocated using different APIs (\texttt{cudaMalloc} vs \texttt{numa\_alloc\_onnode}).}
\label{cudamemcpy}
\end{figure}

As an illustrative example of how different physical allocations, allocation APIs, and data movement APIs interact together in complex ways we evaluate the bandwidth achieved by \texttt{cudaMemcpy} on different types of memory, and show the results in Figure \ref{cudamemcpy}. As \texttt{cudaMemcpy} employs different implementations based on the type of source and destination memory, the test can validate its optimality. The results shown ignore the first warmup run and perform repeated iterations. For this reason, memory allocated through \texttt{cudaMallocManaged} will always have optimal placement. Pinned memory is allocated on the GH200 local to the GPU. We tested also \texttt{cudaMemcpyAsync}, which showed comparable performance. Our results show that the slower host-based implementation is used for all system allocated memory regardless of physical placement.

\section{Microbenchmarks}
We motivate, describe, and show the results of our memory movement-oriented microbenchmarks. We provide a thorough analysis of the bandwidth and latency of memory operations issued by different PUs on the different main memories of the system.

\subsection{Datapaths}
A tightly coupled system like the Quad GH200 is composed of many parts communicating through interconnects with different characteristics and involves different hardware and software subsystems.

\begin{figure}[H]
\centering
\includegraphics[width=\linewidth]{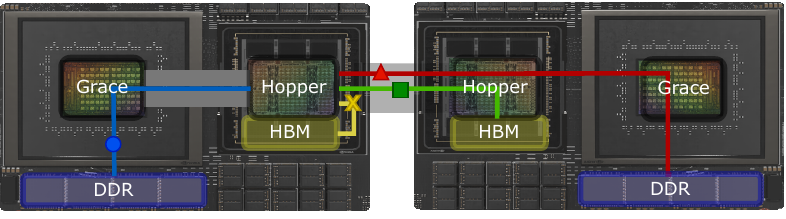}

\includegraphics[width=.5\linewidth]{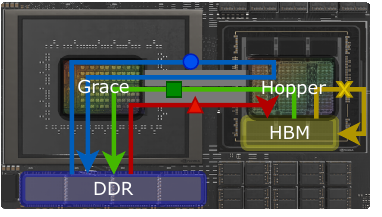}
\caption{Visual representation of the datapaths of different read and write (top) and copy (bottom) operations on the tested system performed by Hopper.}
\label{hopper_grace_transfer}
\end{figure}

Taking operations issued by Hopper as an example, we highlight in Figure \ref{hopper_grace_transfer} all read and write (top) as well as local copy (bottom) datapaths, introducing the colors and symbols used in the rest of the paper. Read and write operations to local HBM (\includegraphics[height=.7em]{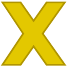}) traverse the shortest path and are limited by the HBM bandwidth of 4000GB/s. All others are limited by the C2C bandwidth of 450GB/s.

Copies require the number of traversals per interconnect to be taken into account. DDR-DDR (\includegraphics[height=.7em]{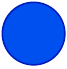}) and HBM-HBM (\includegraphics[height=.7em]{images/x.pdf}) copies traverse the same interconnects twice. For this reason, DDR-DDR copies (\includegraphics[height=.7em]{images/circle.pdf}) are limited to a theoretical bandwidth of half the bandwidth of the interconnect, at 250GB/s. This is lower than the maximum bandwidth achievable by DDR-HBM copies (\includegraphics[height=.7em]{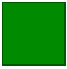} and \includegraphics[height=.7em]{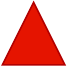}) at 450 GB/s, where every interconnect is traversed only once.

\subsection{Methodology}
We call \textit{kernel} every function that performs a benchmark, excluding all synchronization and measurement code. We develop kernels for both CPU using C++ and inline ARM assembly, and GPU using CUDA. These kernels perform simple memory operations on buffers allocated in different ways and passed as input arguments. Unless stated otherwise, all showed numbers are an average over 10 measurements, discarding the first warmup run.

\subsubsection{Timers}
CPU times are obtained by reading the \texttt{cntvct\_el0} register, a virtual counter that is globally available and uniform across all cores and all GI interconnected GH200. Its frequency can be queried by reading the \texttt{cntfrq\_el0} register and is 1 GHz on our system.
We observe a clock resolution of 32 ns by issuing subsequent reads to the timer.

CUDA supports two on-device timers. The \texttt{\%clock} register can be queried with the \texttt{clock} and \texttt{clock64} library calls. Reading it on different SMs will produce different values. It advances at the device clock speed, which can be queried with \texttt{cudaGetDeviceProperties} and is 1.98 GHz on our system. We observe a clock resolution of 7 cycles or 3.54 ns. The \texttt{\%globaltimer} register can be queried by writing explicit PTX instructions using inline assembly. To find its clock speed we run an experiment consisting of querying the \texttt{\%clock} and the \texttt{\%globaltimer} registers in succession at intervals, and observing the difference increase we find that that the frequency of \texttt{\%globaltimer} is of 1 GHz. We observe a clock resolution of 32 ns, the same for the \texttt{cntfrq\_el0} system timer. The two timers are not directly comparable as they yield unrelated values.

\subsubsection{Multithreaded Benchmarks}
Our test infrastructure allows us to control how many computational resources are used in a kernel. For Grace, we spawn and pin threads to separate cores on application startup. In multi-threaded benchmarks, buffers are equally divided among the threads, such that each one of them works on an equal number of non-overlapping sequential cache lines. A control thread is responsible for choosing a start time step for the test, which is selected by reading the clock from the \texttt{cntfrq\_el0} register and incrementing it by a fixed amount. The control thread communicates to all threads the kernel to run, its arguments, and the chosen start time step. All threads read the clock until the start time step is reached, execute the test, and record the final time step. The total time is taken as the maximum among all final time steps minus the initial time step.

The test setup on Hopper makes use of the cooperative threads API to enable grid-level synchronization. This restricts the grid and block size to be limited such that all threads are active at once, and the kernel must be launched using \texttt{cudaLaunchCooperativeKernel}. A thread selects a starting time step in a way that is analogous to what happens on the CPU by reading the \texttt{\%globaltimer} register and communicates it to all other threads by writing it to global memory. Our tests show that a call to \texttt{\_\_syncthreads} is necessary after grid synchronization for all threads to be aligned and correctly spin waiting for the correct time step before starting the test. As global GPU memory is optimized for coalesced access, all threads access memory in a strided fashion, with a stride equal to the number of threads in the grid, and an initial offset equal to the id of the thread. The end timestep is recorded in shared memory for all threads, and the maximum is taken.

\subsection{Read and Write}
The CPU write kernel uses \texttt{STP} to store 16 bytes using a single instruction. The GPU write kernel stores 8 bytes at a time. Different techniques are used to program the read-only kernel. On CPU, the \texttt{LDP} ARM assembly instruction is explicitly issued. \texttt{LDP} loads two doublewords from memory and stores them into two registers, effectively moving 16 bytes with a single instruction. CUDA kernels cannot rely on this method, as inline \texttt{PTX} instruction can still be optimized out by successive compiler passes. Instead, dummy work is performed at the thread level, in the form of a \texttt{XOR} operation on the read value. Issuing a sufficient number of read operations per cycle is fundamental in achieving peak bandwidth for kernel launches with few threads. We find that using the \texttt{ulonglong2} datatype achieves the best bandwidth for launches with one block of 1024 threads.

\begin{figure}
\centering
\includegraphics[width=\linewidth]{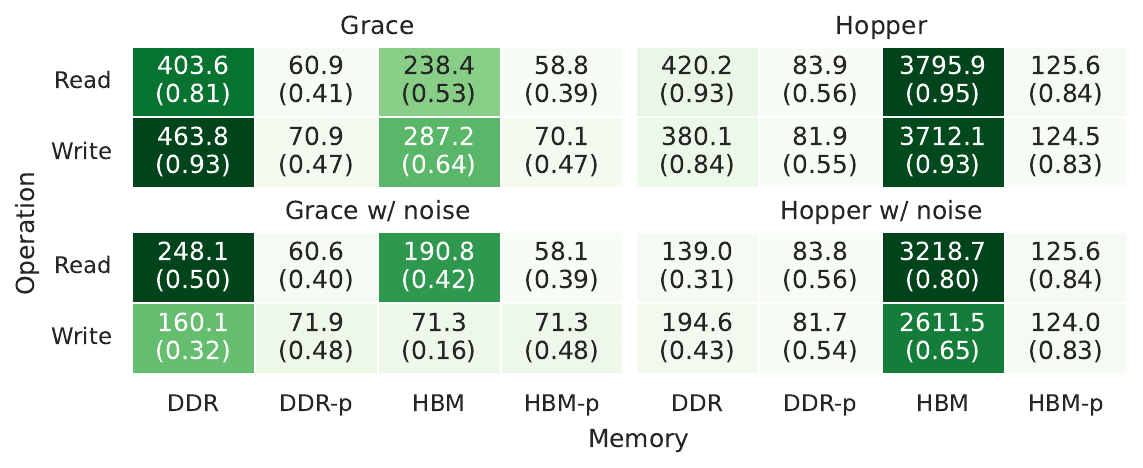}
\caption{Throughput achieved by Grace and Hopper on read and write operations on different types of memory in idle conditions (top) and with load on the C2C interconnect (bottom).}
\label{read_write}
\end{figure}

\begin{figure}
\centering
\includegraphics[width=\linewidth]{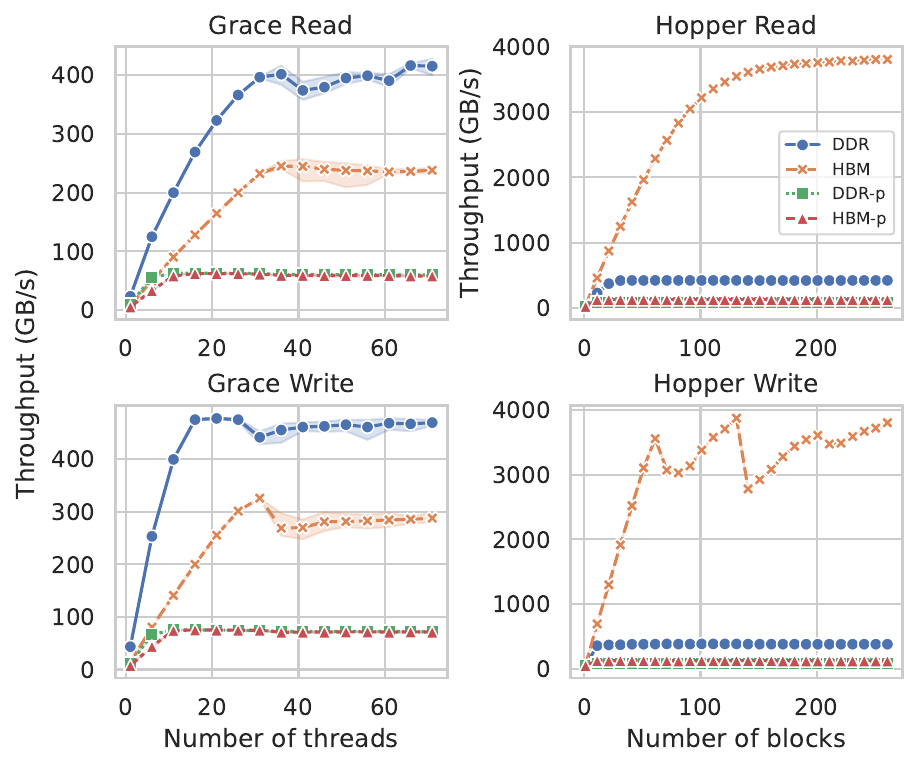}
\caption{Throughput (GB/s) achieved by Grace and Hopper on read and write operations on different types of memory, with different numbers of threads and blocks. Every block is composed of 1024 threads.}
\label{read_write_scalability}
\end{figure}

We also want to measure the bidirectional bandwidth of the C2C interconnect as well as the ability of the system to handle a large number of memory operations issued by multiple PUs on the same GH200. To do this, we develop a Grace and a Hopper \textit{noise kernel} that continuously reads from a large buffer of 8 GB. To stress the C2C interconnect, the Grace noise kernel reads HBM system allocated memory and the Hopper noise kernel reads DDR allocated memory. We start the noise kernel for one PU and run the read and write tests for the other PU.

We report our results in Figure \ref{read_write}. We show the achieved bandwidth in GB/s as well as the ratio of achieved bandwidth over maximum theoretical bandwidth according to Figure \ref{bounds}. In the simple benchmarks, Hopper is better at making use of the C2C interconnect when accessing local memory compared to Grace, with read and write bandwidth to DDR of 93\% and 84\% respectively, compared to the 53\% and 64\% achieved by Grace in operation to HBM. Operations that cross both the C2C interconnect and NVLink incur considerable overheads and never go above 60\% of theoretical bandwidth.

When adding noise, accesses to peer GH200 memory are not affected. Bandwidth to local DDR is limited by the bandwidth of 500 GB/s that needs to be split between two PUs. Writes to HBM are the most impacted, with a Grace bandwidth of 17\% and a Hopper bandwidth of 65\% of the theoretical maximum. Summing the bandwidth of both PUs 2682.8 GB/s are reached, which is only 67\% of the theoretical maximum.

\subsection{Copy}
\begin{figure}
\centering
\includegraphics[width=\linewidth]{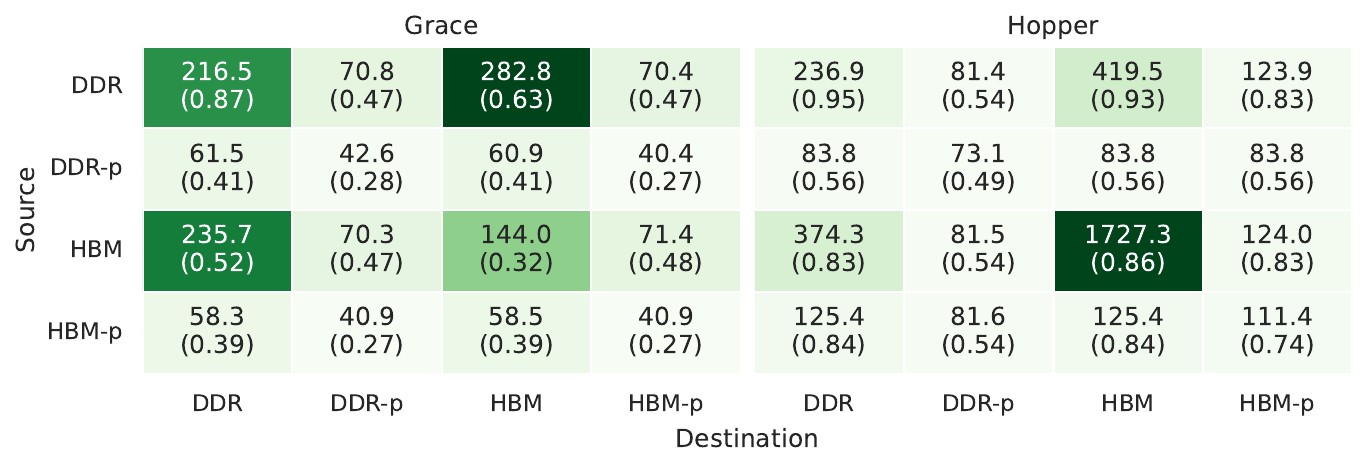}
\caption{Copy throughput (GB/s) achieved by Grace and Hopper on different source and destination memory types.}
\label{global_copy}
\end{figure}
\begin{figure}
\centering
\includegraphics[width=\linewidth]{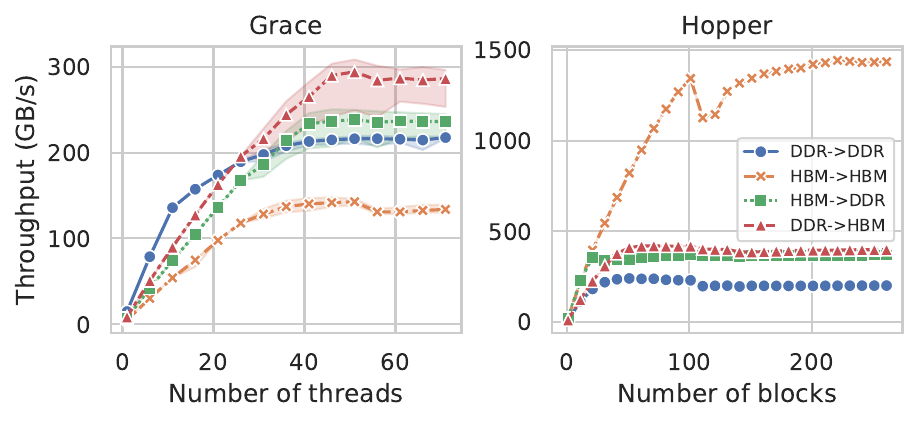}
\caption{Throughput achieved by Grace and Hopper on intra-GH200 copy operations on different types of memories, with different numbers of threads and blocks. Every block is composed of 1024
threads.}
\label{copy_scalability}
\end{figure}

In the Grace copy kernel, a single loop iteration contains four pairs of \texttt{LDP} and \texttt{STP} instructions on separate pairs of registers to ensure pipelining. The Hopper kernel performs 8-byte wide copies in a stride equal to the number of threads. Throughput is measured as the size of the buffer over transfer time. The benchmarks highlight the duplex characteristics of some interconnects on the system. For example, copying from DDR to DDR results in a measured bandwidth that is about half the bandwidth of a read operation.

Results for both Grace and Hopper are shown in Figure \ref{global_copy}. We note the following behaviors:
\begin{itemize}
    \item There are asymmetries in memory transfers. Grace achieves a higher throughput when copying from local memory to a peer GH200 compared to the opposite direction. Local DDR to HBM transfers are faster than HBM to DDR transfers.
    \item Hopper does a better job at utilizing the available bandwidth when crossing multiple interconnects.
\end{itemize}

Figure \ref{copy_scalability} shows the scalability of the copies performed by Grace and Hopper for the cases of system-allocated memories that reside on a single GH200.

\subsection{Latency Benchmarks}
We measure the latency of main memory accesses and of core-to-core communication.
\subsubsection{Pointer Chase}
\begin{figure}
\includegraphics[width=\linewidth]{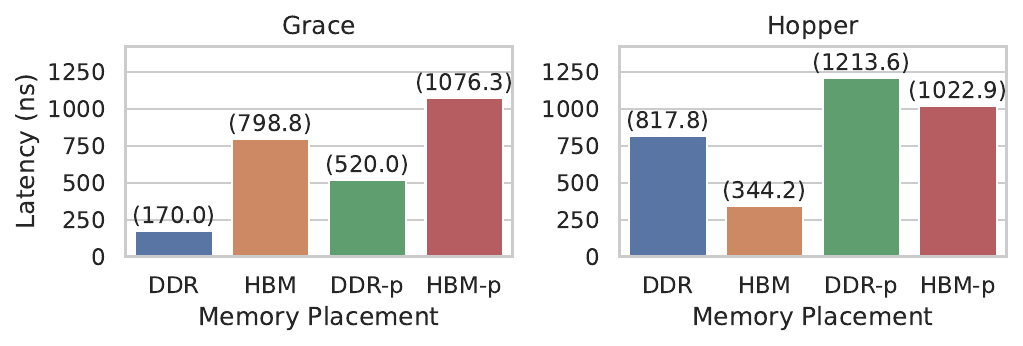}
\caption{Main memory access latency of Grace and Hopper.}
\label{latency}
\end{figure}

\begin{figure*}
\centering
\includegraphics[width=\linewidth]{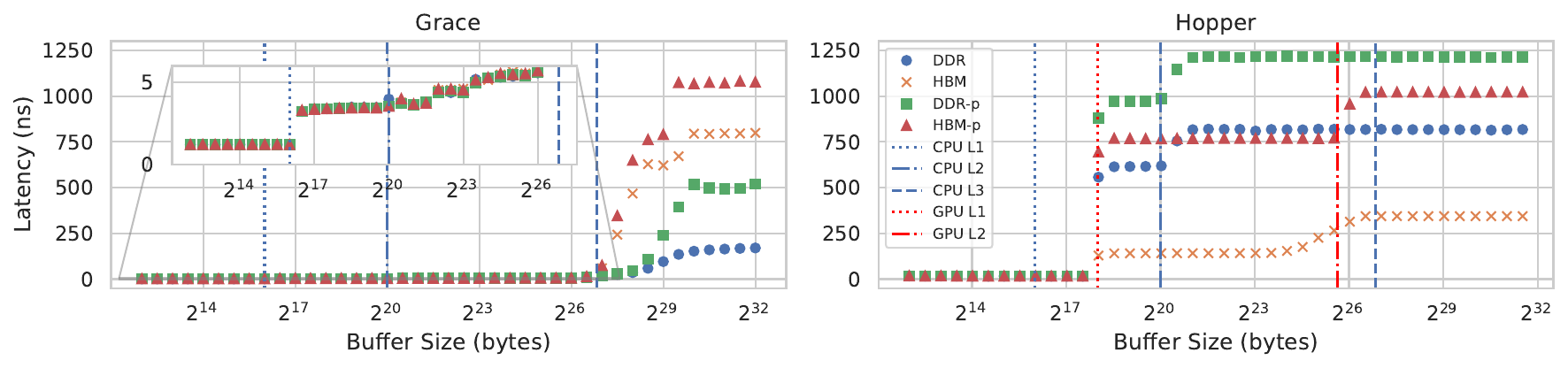}
\caption{Latency of accesses to different types of memory, with varying buffer sizes. Results are shown for both Grace and Hopper. The plots show the size of the relevant caches on the system.}
\label{latency_scalability}
\end{figure*}

To measure the access latency to memory we employ a pointer chase benchmark. We modify Google's multichase benchmark \cite{multichase} to support NUMA allocations and a pointer chase GPU kernel. The benchmark performs a pointer chase for 2.5 seconds, recording the number of accesses with a granularity of 200. We show memory access latency in Figure \ref{latency}. Accesses that cross the C2C interconnect (Grace to HBM and Hopper to DDR) show the same latency. The same behavior is also shown by access to peer HBM. We show the scalability of memory access latency on increasing buffer size in Figure \ref{latency_scalability}. As the pointer chase iterates on the same buffer multiple times, we can also measure the latency of accesses to the cache. Cache sizes are highlighted using vertical lines. Hopper displays a simple caching behavior, with all types of memories showing the same latency if the buffer size is within some cache bound. Hopper, on the other hand, shows a behavior that is dependent on where the memory is physically allocated. We make the following key observations:
\begin{itemize}
    \item If the buffer fits in Hopper L1 cache, the latency is the same regardless of the physical allocation. For buffers larger than L1 the behavior changes for all types of physical memory allocations.
    \item For memory physically allocated on DDR the Grace L2 cache affects latency. For buffers larger than Grace L2 the latency never changes, showing that these accesses are never cached.
    \item Hopper L2 cache can cache data that is physically allocated on HBM, both local and peer. L2 resident peer HBM accesses are faster than local DDR accesses.
\end{itemize}
This behavior highlights great differences between how cache coherency and memory accesses are handled between Grace and Hopper.

\subsubsection{Ping-Pong}
\begin{figure}[H]
\includegraphics[width=\linewidth]{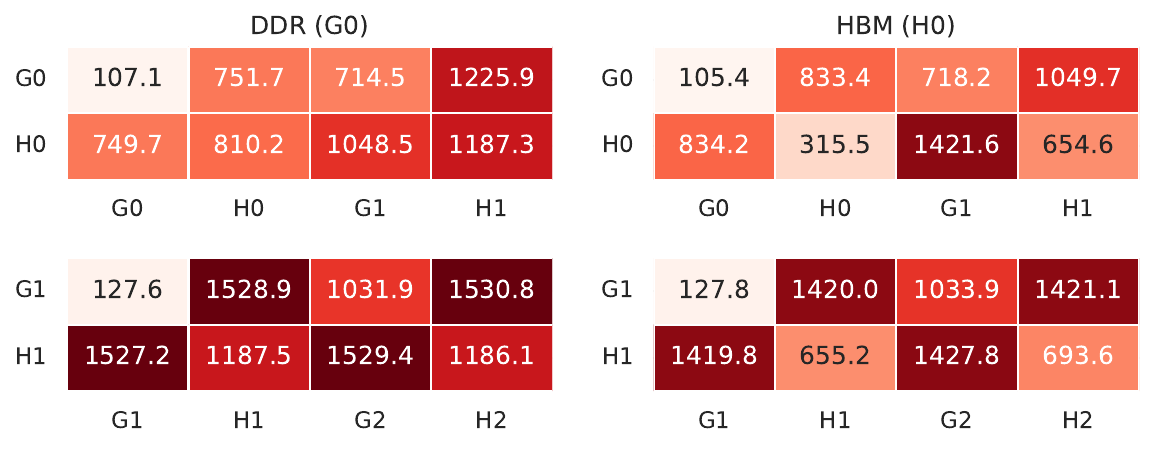}
\caption{Latency in nanoseconds of a full ping-pong exchange between different combinations of PUs. G and H stand respectively for Grace and Hopper. The number refers to the Superchip unit the PU belongs to. All memory allocations are done on the GH-0. Hopper-Hopper local communications are between different SMs.}
\label{pingpong}
\end{figure}

We run a ping-pong benchmark to evaluate the communication latency inside a PU and between different PUs. This benchmark highlights the characteristics of the cache coherent interconnects, as well as the behavior of the cache coherency protocol when dealing with atomics.
We leverage the atomic compare-and-swap (CAS) operation and the interoperability of atomic types between host and device offered by CUDA with the \texttt{cuda::std::atomic} type. CAS can be used as an atomic conditional store, writing a \textit{desired} value to a memory location if it contains an \textit{expected} value. To avoid any spurious contention, an atomic flag of one byte is placed in a chunk of memory of the size of two cache lines. The atomic flag is initialized with a PONG value.
Two threads, one running the ping function and the other running the pong function, are started on the PUs of interest. The ping function and the pong function have an implementation both for the host and for the device. The ping function conditionally sets the flag to PING if its value is PONG, while the pong function does the opposite. Time is measured since the first successful communication. The ping pong host functions are compiled using GCC 13.2 with support for Neoverse-v2 to enable the issue of the ARM CASB instruction.%(a CAS operating on one byte), the host functions are compiled separately using GCC-13, which is the only version to fully support the Neoverse-v2 architecture, but is not yet supported by the latest CUDA version at the time of writing.

Core-to-core tests show non-uniform latencies depending on where the memory is physically allocated. This is a result of the cache coherency protocol and the way that atomic operations are requested and executed. Figure \ref{pingpong} shows the results of the ping-pong benchmark being run between different PUs with different allocations of the flag. All allocations are done on GH-0. Exchanges involving the GH200 where the memory is physically allocated are faster, with local exchanges being the fastest. Hopper-Hopper communications benefit greatly from memory being physically allocated on HBM. All Grace-Hopper communications benefit from memory being allocated on the HBM of the Hopper participating in the communication, except for local Grace ping to remote Hopper pong. Grace-Grace communication is faster if the memory is allocated on a participating GH, but shows no difference based on the type of memory.

\subsection{Internode Benchmarks}
We evaluate the bandwidth of internode communications in the Alps supercomputer. Our benchmarks are based on MPI, the dominant programming model for parallel and distributed architectures.
We use Cray MPICH 8.1.28 on top of Libfabric 1.15.2. We activate the GPU Transport Layer (GTL) to handle buffers allocated with \texttt{cudaMalloc} directly using DMA.

\begin{figure}
\centering
\includegraphics[width=.9\linewidth]{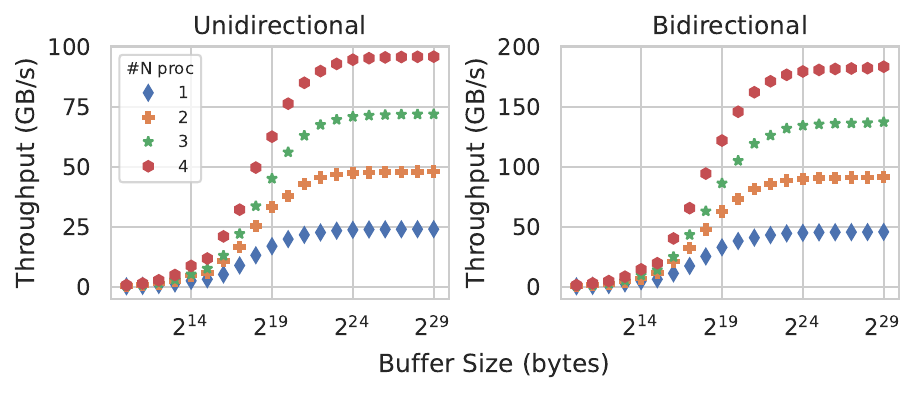}
\caption{Throughput scalability of memory transfers using \texttt{MPI\_Isend} and \texttt{MPI\_Irecv} for different numbers of processes per node.}
\label{mpi_send_recv_scalability}
\end{figure}

We run a node-to-node network bandwidth test using \texttt{MPI\_Isend} and \texttt{MPI\_Irecv}. In Figure \ref{mpi_send_recv_scalability} we show the results scaling both buffer size and the number of processes per node, in both the unidirectional case and the bidirectional case. As an MPI process can utilize only one network interface, four processes are required to utilize the full bandwidth of a node. The amount of data to be transferred is equally split among the processes of the node. Processes are assigned to different NUMA nodes. We show the results for allocations on local DDR only as we find negligible differences when using other types of allocations.

Only one of the two nodes performs the measurement. Analogous to our multithreaded benchmarks, one of the processes of the measuring nodes selects a starting time and communicates it to all other local processes. The final time is taken as the maximum among all local processes.

\section{Applications}
We show the performance results of simple applications as a function of the physical memory placement, using the same framework and terminology as our microbenchmarks.

\subsection{GEMM}
\begin{figure}
\centering
\includegraphics[width=\linewidth]{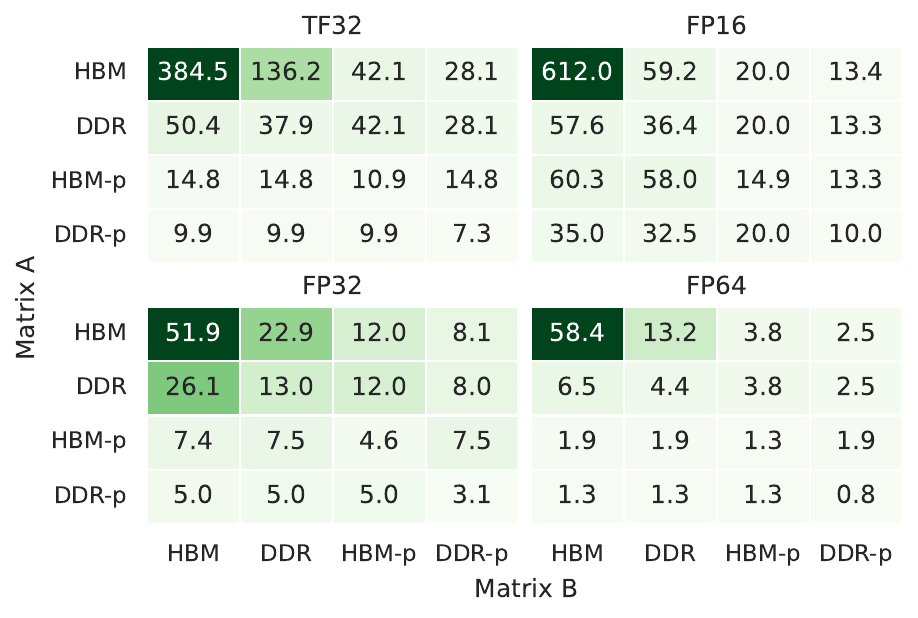}
\caption{Performance in TFLOPS, of GEMM operations performed on square matrices placed in different memory locations.}
\label{gemm}
\end{figure}

\begin{figure}
\centering
\includegraphics[width=\linewidth]{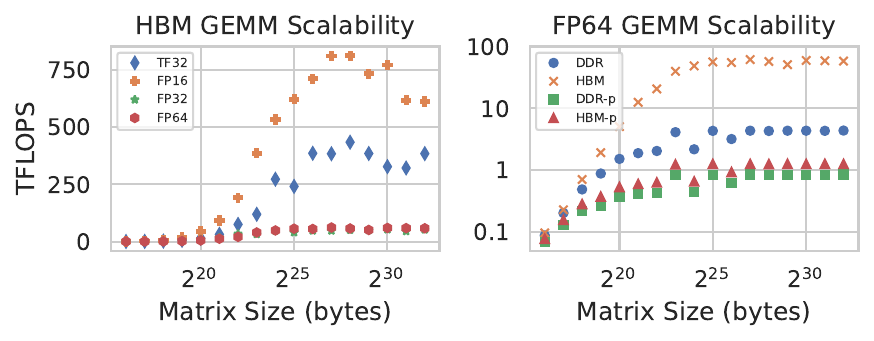}
\caption{Performance scalability of GEMM operation with varying matrix sizes. In the left plot, all data is physically allocated in local HBM and the datatype changes. In the right plot, all operations are done using FP64, and the physical allocation of source matrices changes.}
\label{gemm_scalability}
\end{figure}

\begin{table}
\centering
\caption{Advertised TFLOPS for different datatypes on the H100 SXM GPU, and if they leverage Tensor Cores.}
\begin{tabular}{c c c }
    \toprule 
    \textbf{Type} & \textbf{Tensor Core} & \textbf{TFLOPS} \\
    \midrule
    FP16 & \ding{51} & 989 \\
    TF32 & \ding{51} & 494 \\
    FP32 & \ding{55} & 67 \\
    FP64 & \ding{51} & 67 \\
    \bottomrule
\end{tabular}
\label{tflops}
\end{table}

Matrix-matrix multiplication is a fundamental building block of scientific computing applications. Its importance led to the development of the Tensor Core, a dedicated multiply-and-accumulate unit on NVIDIA GPUs \cite{markidis2018nvidia}. The rise of Large Language Models (LLMs) and the transformer architecture \cite{vaswani2017attention} has made the performance of this operation critical for machine learning workloads as well.

Figure \ref{gemm} shows the performance of a GEMM operation multiplying source matrices $A$ and $B$. Our tests show that reads have the greatest impact, while the placement of the destination matrix has a negligible effect on performance. We show data for experiments where the destination matrix C is always placed in local HBM memory. The implementation is provided by cuBLAS 12.3 and uses a single Hopper. We measure the performance achieved using different data types. We report in Table \ref{tflops} the advertised TFLOPS for the H100 SXM GPU for different datatypes, which are analogous to the Hopper in GH200. Every matrix is 4 GB. We find that this size is enough to hide the effect of caches and reaches asymptotic throughput.

We also measure the performance scalability with increasing matrix size and show the results in Figure \ref{gemm_scalability} for fixed physical memory allocation and varying datatype, as well as fixed datatype and varying the physical memory allocation. We round down the size of the matrix to be a multiple of 8, to make sure that the tensor cores are fully utilized.

We observe that except for FP16, where smaller sizes that fit in the cache see a higher throughput, HBM provides enough bandwidth to make the workload compute-bound. As soon as one of the matrices is moved away from HBM, the performance drops and the workload becomes heavily memory-bound, especially for the datatypes making use of Tensor Cores. The access patterns for matrices $A$ and $B$ are different. This is reflected in the asymmetry of the matrices in figure \ref{gemm}.

\subsection{LLM inference}
\begin{figure}
\centering
\includegraphics[width=\linewidth]{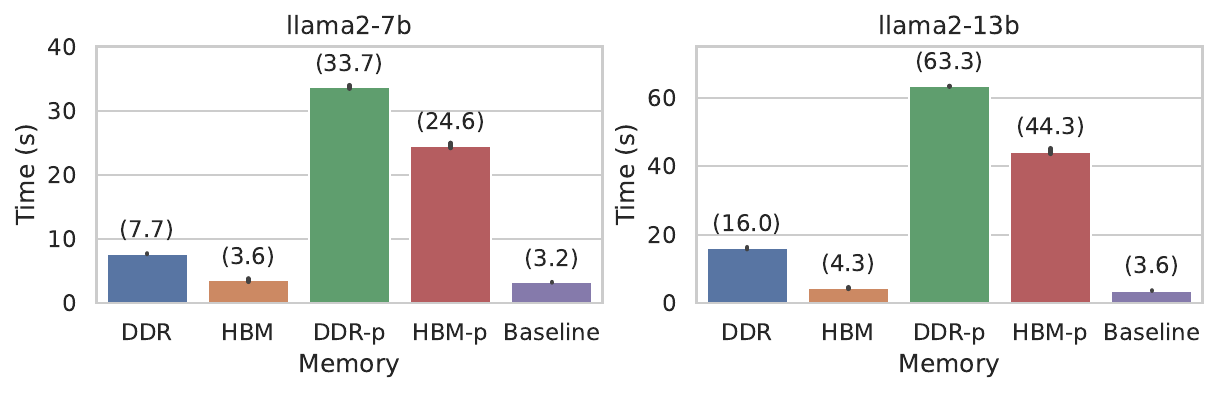}
\caption{LLM inference time for 100 tokens, for different models, with different physical memory allocations (lower is better).}
\label{llama}
\end{figure}

The growth in the size of LLMs has led to memory footprint becoming a fundamental problem in their training and deployment \cite{hoefler2021sparsity,frantar2022gptq}. Having access to a larger pool of memory opens up opportunities to run these workloads using fewer machines.

We run an inference workload in the Llama2-7b and Llama2-13b models \cite{touvron2023llama} using the HuggingFace APIs to generate 100 tokens from an empty prompt, using the \texttt{torch.float16} datatype, using PyTorch 2.2.0 \cite{paszke2019pytorch} and Python 3.11.7. We use the pluggable allocator functionality of the PyTorch library to control memory allocations. Due to stability and performance issues, small allocations (less than 1 MB) are done using \texttt{cudaMallocAsync}, while large allocations are done using \texttt{numa\_alloc\_onnode}. The peak memory utilization for small and large allocations is respectively 133 MB and 27 GB for Llama2-7b, and 168 MB and 52 GB for Llama-13b, making small allocations negligible ($<1\%$).

We show results in Figure \ref{llama}. Memory access speed plays a fundamental role in the throughput. Compared to purely memory-bound synthetic workloads, however, the difference in performance is less dramatic. We also show the baseline performance. Our allocator is slower as it incurs synchronization overheads for large allocations.

\subsection{NCCL}
\begin{figure}
\centering
\includegraphics[width=\linewidth]{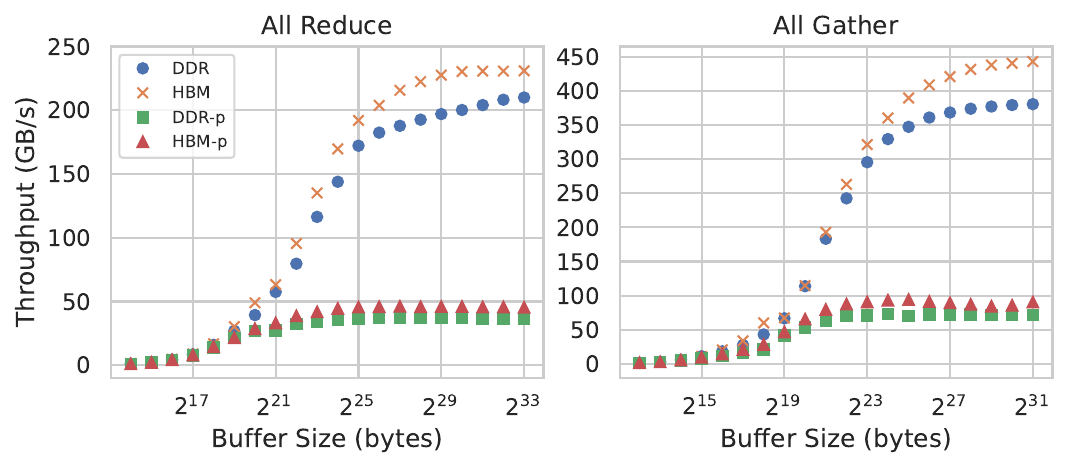}
\caption{Performance scalability of intra-node all reduce and all gather NCCL operations performed by four processes using four GPUs.}
\label{nccl_a2a}
\end{figure}
\begin{figure}
\centering
\includegraphics[width=\linewidth]{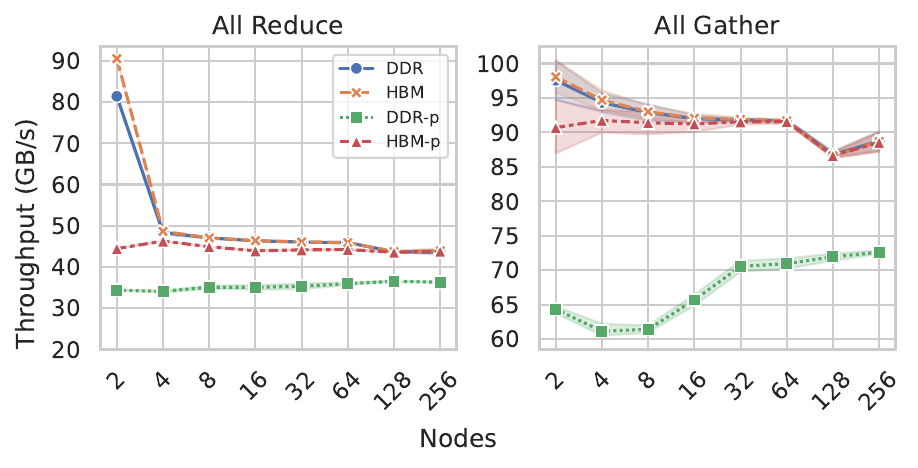}
\caption{Performance scalability of inter-node all reduce and all gather NCCL operations. Every node runs four processes, each using a single GPU.}
\label{nccl_a2a_scalability}
\vspace{-1em}
\end{figure}
NVIDIA Collectives Communication Library (NCCL) is a host library that implements various communication primitives for NVIDIA GPUs. It supports multi-GPU setups, both single-node and multi-node, and makes use of PCIe, NVLink, and networking transparently. It provides the building blocks necessary to develop large-scale multi-GPU applications.

We show our results for the all reduce and all gather operations. Bandwidth is calculated as the size of the buffer over the time it takes to complete the operation. In our tests system allocated memory on HBM and memory allocated through \texttt{cudaMalloc} showed the same performance. In Figure \ref{nccl_a2a} we show the performance scalability with increasing buffer size when running four processes on the same node. Our results show the importance of locality, with same-GH200 memory greatly outperforming peer access, and HBM and DDR showing similar throughput.

In Figure \ref{nccl_a2a_scalability} we show the performance scalability with an increasing number of nodes participating in the collective, with four processes per node. The size of the buffer for the all-reduce operation is 4 GB, while the size of the buffer for the all-gather operation is 16 MB times the number of processes participating in the collective. Peer DDR memory access severely limits the performance of the collectives. In all other cases, performance differences are negligible.

Our results show that Superchip locality, more than the type of memory used, plays an important role in applications making heavy use of collective operations across multiple processes.
\section{Conclusions}
This paper offers a comprehensive view of the memory hierarchy within the Quad GH200 node configuration of the Alps supercomputer. We conduct benchmarks on read, write, and copy operations across all combinations of physical memory allocations and processing units. Our analysis relates the measured performance to the theoretical bounds provided by the datapaths of the individual operations. Additionally, we present performance figures for example applications, highlighting the significance of data placement and memory access patterns for memory-bound workloads.

We argue that despite the sophisticated memory system of the Quad GH200 node, looking at the system in terms of individual interconnected Superchips is crucial to achieving good performance. The C2C interconnect lives up to its promise and opens up possibilities for the development of heterogeneous applications mixing CPU and GPU computations, and for effectively extending the pool of memory available to PUs.

\bibliographystyle{ieeetr}
\bibliography{references}

\vspace{12pt}

\end{document}